\documentclass[10pt]{article}
\usepackage{float}
\usepackage[symbol]{footmisc} 
\usepackage[superscript,sort]{cite}
\usepackage{array}
\usepackage{bm}

\newcolumntype{x}[1]{%
>{\centering\hspace{0pt}}p{#1}}%
\usepackage[normalem]{ulem} 
\usepackage[colorlinks=true,linkcolor=black,citecolor=black,urlcolor=black,bookmarks=true,breaklinks=true]{hyperref}
\usepackage[usenames]{color}
\usepackage{amsmath,amssymb}

\usepackage{amsthm,amscd,amsxtra,amsfonts,amsmath,amssymb,multirow}
\usepackage{wrapfig}
\usepackage[footnotesize]{caption}
\usepackage{subcaption}
\usepackage[tiny,compact]{titlesec}
\usepackage{threeparttable} 
\usepackage{helvet}

\usepackage{booktabs}
\usepackage[table]{xcolor}
\usepackage{xcolor,colortbl}
\usepackage{tikz}
\usetikzlibrary{shapes,arrows}
\usepackage[T1]{fontenc}
\usepackage[linesnumbered,ruled]{algorithm2e} 
\titlespacing*{\section}{0pt}{*0}{*0}
\titlespacing*{\subsection}{0pt}{*0}{*0}
\titlespacing*{\subsubsection}{0pt}{*0}{*0} 
\titlespacing{\paragraph}{0pt}{*0}{*1}

\definecolor{MyPurple}{rgb}{1,0,1}

\newcommand{\beq}[1]{\begin{equation} \label{#1}}
\newcommand{\eeq}{\end{equation}}
\newcommand{\barray}{\begin{array}{ll}}
\newcommand{\earray}{\end{array}}

\newcommand\BindPCCBest{0.828}
\newcommand\BindRMSEBest{1.37}
\newcommand\BindPCCMedian{0.826}
\newcommand\BindRMSEMedian{1.37}
\newcommand\MutTestPCCMedian{0.74} 
\newcommand\MutTestRMSEMedian{1.07}
\newcommand\MutTestSeqPCCMedian{0.81} 
\newcommand\MutTestSeqRMSEMedian{0.94}
\newcommand\MutCVPCCMedian{0.71}
\newcommand\MutCVRMSEMedian{1.06}
\newcommand\MutCVSeqPCCMedian{0.77}
\newcommand\MutCVSeqRMSEMedian{0.94}
\newcommand\MemMTLPCCMedian{0.52} 
\newcommand\MemMTLRMSEMedian{1.07}
\newcommand\MemSinglePCCMedian{0.48} 
\newcommand\MemSingleRMSEMedian{1.20}

\begin{document}
\pagenumbering{roman}

\clearpage \pagebreak \setcounter{page}{1}
\renewcommand{\thepage}{{\arabic{page}}}

\title{TopologyNet: Topology based deep convolutional neural networks for biomolecular property predictions
}

\author{
Zixuan Cang$^1$,
 and
Guo-Wei Wei$^{1,2,3}$ \footnote{ Address correspondences  to Guo-Wei Wei. E-mail:wei@math.msu.edu}\\
$^1$ Department of Mathematics \\
Michigan State University, MI 48824, USA\\
$^2$  Department of Biochemistry and Molecular Biology\\
Michigan State University, MI 48824, USA \\
$^3$ Department of Electrical and Computer Engineering \\
Michigan State University, MI 48824, USA \\
}

\date{}
\maketitle
\maketitle
\abstract{
Although deep learning approaches have had tremendous success in image, video  and audio processing,  computer vision, and  speech recognition,
their applications to three-dimensional (3D) biomolecular structural data sets have been hindered by the entangled geometric  complexity and biological complexity.  We introduce topology, i.e., element specific persistent homology (ESPH), to untangle geometric  complexity and biological complexity. ESPH represents 3D complex geometry by one-dimensional (1D) topological invariants  and retains crucial biological information via a multichannel image representation. It is able to reveal hidden structure-function relationships in biomolecules. We further integrate ESPH and convolutional neural networks to construct a multichannel topological neural network (TopologyNet) for the predictions of protein-ligand binding affinities and protein stability changes upon mutation. To overcome the limitations to deep learning arising from small and noisy training sets, we present a multitask topological convolutional neural network (MT-TCNN). We demonstrate that the present TopologyNet architectures  outperform other state-of-the-art methods in the predictions of protein-ligand binding affinities, globular protein mutation impacts and membrane protein mutation impacts.}

\section{Introduction}
The understanding  of the structure-function relationships of biomolecules, such as the prediction of protein-ligand binding affinity and protein stability change upon mutation from three-dimensional (3D)  structures,   is the holy grail of computational biophysics and a  central issue in experimental biology.  Numerous approaches have been developed to unveil these relationships. 
Physics based models make use of fundamental laws of physics, i.e., quantum mechanics (QM)  \cite{Warshel:1976, Cui:2002, YZhang:2009a}, molecular mechanics (MM)  \cite{McCammon:1977,CHARMM22, AMBER15}, continuum mechanics \cite{Roux:1999,Warshel:1998,Sharp:1990a,Tully-Smith:1970}, 
multiscale modeling  \cite{Holst:1993,Baker:2004,Dong:2008MCB}, 
statistical mechanics, thermodynamics,  etc, to both understand and predict  structure-function relationships. These approaches provide physical insights and are  indispensable for our basic understanding of the relationship between protein structure and function. 

The exponential growth of biological data has set the stage for data-driven  discovery of structure-function relationships. Indeed,  the Protein Data Bank (PDB) has   accumulated  more than 125,000  tertiary structures.  The availability of these 3D structural data  enables knowledge based approaches to offer  complementary and  competitive predictions of  structure-function relationships.  
 The recent advances in machine learning algorithms  have made data driven approaches more  competitive and powerful than ever. Arguably, machine learning is one of the most important developments in data analysis.  
Machine learning has become an indispensable tool in biomoelcular data analysis and prediction. Virtually every computational problem in computational biology and biophysics, such as the predictions of solvation free energies,  protein-ligand  binding affinities, mutation impacts, pKa values, etc,  has a class of knowledge based approaches that are either parallel or complementary to physics based approaches. 
With its ability to recognize nonlinear and high-order interactions among features as well as the capability of handling data with underlying spatial dimensions, deep convolutional neural networks have led to breakthroughs in image processing, video, audio and computer vision~\cite{krizhevsky2012imagenet,simonyan2014very},    whereas recurrent nets  shed light on sequential data such as text and speech \cite{lecun2015deep,hinton2012deep,schmidhuber2015deep,ngiam2011multimodal}. 
Deep learning has fueled the rapid growth in several areas of data science \cite{schmidhuber2015deep,lecun2015deep,ngiam2011multimodal}.
Machine learning based approaches are advantageous because of their ability to handle very large data sets and nonlinear relationships in physically derived descriptors. In particular, deep learning has the ability to automatically extract optimal features and discover intricate structures in large data sets.
  
When there are multiple learning tasks,   multi-task learning (MTL)~\cite{evgeniou2004regularized,caruana1998multitask} provides a powerful tool to exploit the intrinsic relatedness among learning tasks, transfer predictive information among tasks, and achieve better generalized performance. During the learning stage, MTL algorithms seek to learn a shared representation (e.g., shared distribution of a given hyper-parameter~\cite{evgeniou2004regularized}, shared low-rank subspace~\cite{evgeniou2007multi,pong2010trace}, shared feature subset~\cite{liu2009multi} and clustered task structure~\cite{zhou2011clustered}), and use the shared representation to bridge between tasks and transfer knowledge. 
MTL has found applications to   bioactivity of small molecular drugs  \cite{hughes2015modeling,unterthiner2015toxicity,lusci2013deep,wallach2015atomnet} and genomics \cite{dahl2014multi, ramsundar2015massively}. Linear regression based MTL heavily depends on the well crafted features while neural network based MTL allows more flexible task coupling and is able to deliver decent results with large number of low level features as long as such features have the representation power of the problem.

For complex 3D biomolecular data, the physical features used in machine learning vary greatly in their nature. Typical features are generated from  geometric properties, electrostatics, atomic type, atomic charge and graph theory properties  \cite{BaoWang:2016HPK}. 
Such manually extracted features can be fed to a deep neural network, but the performance heavily relies on the fashion of feature construction. On the other hand, convolutional neural network is able to learn high level representations from low level features. However, the cost is huge for directly applying convolutional neural network to the 3D biomolecules when long range interactions need to be considered.
To the best of our knowledge, there currently is no competitive deep learning algorithm for predicting protein-ligand binding affinities and protein stability changes upon mutation from 3D biomolecular data sets.    Additionally, there is a pressing need to design a robust  multi-task deep learning method for improving both protein-ligand binding affinity and mutation impact predictions.
 A major obstacle in the development of deep learning nets for 3D biomolecular data is their entanglement between 
intrigue geometric complexity and  biological complexity \cite{ZYu:2008, XFeng:2012a,QZheng:2012,NKWH07,JLi:2013}.

 Most theoretical models for the study of structure-function  relationships of biomolecules    are    based  on   geometric modeling techniques 
\cite{JLi:2013,DasGupta2016}. Mathematically, these approaches  make use of local  geometric information,  i.e., coordinates, distances, angles, areas 
and sometimes curvatures \cite{DDNguyen:2016c} for the physical modeling of biomolecular systems.   Indeed, the importance of geometric modeling for structural biology \cite{XFeng:2012a},
and  biophysics \cite{XFeng:2013b,KLXia:2014a,PMach:2011} 
cannot be overemphasized. However, geometry based models are often inundated with too much structural detail and  are frequently computationally intractable. In many biological problems, such as the  opening or closing of ion channels, the association or disassociation of binding ligands,  the folding or unfolding of proteins, the symmetry breaking or formation of virus capsids \cite{Twarock:2008},  there exist  obvious topological changes. In fact, one only needs qualitative topological information, not quantitative, to understand many physical and biological functions. Put another way,  in many biomolecular systems there are {\it topology-function relationships}.

Topology presents entirely different approaches and could provide dramatic simplification to biomolecular data \cite{Schlick:1992trefoil,Zomorodian:2005,sumners:1992,IKDarcy:2013,CHeitsch:2014,Demerdash:2009,DasGupta2016,XShi:2011}. 
The study of topology   deals with the connectivity of different components in a space, and  characterizes independent entities,  rings and higher dimensional faces within the space \cite{kaczynski:mischaikow:mrozek:04}. Topological methods provide the ultimate level of abstraction of many biological processes, such as the  open or close state of ion channels, the assembly or disassembly of virus capsids, the folding and unfolding of proteins, and the association or disassociation of ligands. 
The fundamental task of topological data analysis is to extract topological invariants, namely the intrinsic features of the underlying space, of a given data set without additional structure information. Examples include covalent bonds, hydrogen bonds, van der Waals interactions, etc.
A fundamental  concept in  algebraic topology is simplicial homology, which concerns the identification of topological invariants from a set of discrete node coordinates such as atomic coordinates in a protein  or a protein-ligand complex. For a given (protein)  configuration, independent components, rings and cavities are topological invariants and their numbers  are called Betti-0, Betti-1 and Betti-2, respectively.
However, conventional topology or homology is truly free of metrics or coordinates, and thus retains too little geometric information to be practically useful. 

Persistent homology   is a relatively new branch of algebraic topology that   embeds multiscale geometric information into topological invariants to achieve an  interplay between geometry and topology. It creates a variety of topologies of a given object by varying a  filtration parameter, such as the radius of a ball or the level set of a surface function. 
As a result, persistent homology can capture topological structures continuously over a range of spatial  scales. Unlike commonly used computational homology which results in  truly metric free  representations, persistent homology embeds geometric information in topological invariants, e.g., Betti numbers 
so that ``birth"  and ``death" of  isolated components, circles, rings,  voids or cavities can be monitored at all geometric scales by topological measurements.
In the past decade, persistent homology has been developed as a new multiscale representation of topological features.  
The 0-th dimensional version was originally introduced for computer vision applications under the name ``size function" \cite{Fro90, Frosini:1999,Robins:1999}.  
Persistent homology theory and a resulting algorithm was formulated  by Edelsbrunner et al. \cite{Edelsbrunner:2002}. Later, a more general theory was developed by Zomorodian and Carlsson \cite{Zomorodian:2005}. 
Since that time, there has been significant theoretical development \cite{BH11,CEH07,CEH09,CEHM09,CCG09,CGOS11,Carlsson:2009theory,CSM09,SMV11,zigzag}, as well as various computational algorithms \cite{OS13,DFW14,Mischaikow:2013,javaPlex,Perseus, Dipha}. 
Persistent homology is often visualized by the use of barcodes \cite{CZOG05,Ghrist:2008} where horizontal line segments or bars represent homology generators that survive over different filtration scales.

Persistent homology  have been applied to computational biology \cite{Kasson:2007,Gameiro:2014,Dabaghian:2012, Perea:2015b}, 
 such as  mathematical modeling and prediction of  nano particles, proteins and other biomolecules \cite{KLXia:2014c, KLXia:2015a,Gameiro:2014}. We have introduced molecular topological fingerprint (TF)   to reveal topology-function relationships in protein folding and protein flexibility \cite{KLXia:2014c}. We demonstrated that in the field of biomolecule analysis, contrary to the commonly held belief in many other fields, short-lived topological events are not noisy, but part of TFs. 
Quantitative topological analysis has been developed to predict the curvature energy of fullerene isomers  \cite{KLXia:2015a,BaoWang:2016a} and protein folding stability \cite{KLXia:2014c}.   Differential geometry based persistent homology  \cite{BaoWang:2016a},  
  multidimensional persistence \cite{KLXia:2015c}, and  multiresolutional persistent homology \cite{KLXia:2015e,KLXia:2015d}
	have been proposed to 	better characterize  biomolecular data  \cite{KLXia:2015c},  detect protein cavities \cite{ESES:2017},  
 and resolve ill-posed inverse problems in  cryo-EM  structure determination \cite{KLXia:2015b}. 
	Persistent homology based machine learning algorithm has also been developed for  protein structural classification  \cite{ZXCang:2015}.

However, current persistent homology oversimplifies biological information during the topological simplification of geometric complexity. 
Consequently, persistent homology based machine learning algorithms  were not as competitive as other conventional techniques in protein classification  \cite{ZXCang:2015, kusano2016persistence}. 

The objective of the present work is to introduce a new topology, namely, element specific persistent homology (ESPH), to untangle the geometric complexity and biological complexity in biomolecular data sets and to reveal the hidden structures in biomolecules. We further  develop ESPH based neural network (TopologyNet) models  for the prediction of   biomolecular structure-function relationships.  Specifically, we integrate ESPH and  convolutional neural networks (CNNs) to significantly improve the state-of-the art methods for protein-ligand binding affinity and protein mutation impact predictions from 3D biomolecular data. In this approach, topological invariants are used to reduce the dimensionality of  3D biomolecular data. Additionally, element specific persistent barcodes offer  image-like topological representation to facilitate convolutional deep neural networks. Moreover, biological information is retained by 
element specific topological fingerprints and described as multichannels in our image like representation. Furthermore,  convolutional   neural networks uncover hidden relationships between biomolecular topological invariants and biological functions. Finally, a multitask  topological convolutional   neural network (MT-TCNN) framework is introduced  to exploit the relations among various  structure-function predictions and enhance the prediction for problems with  small and noisy training data. Our hypothesis is that many biomolecular predictions share a common set of topological fingerprints and are highly correlated  to each other. As a result, multitask deep learning based on simultaneous training and prediction will improve upon existing predictions.

\section{Methods}\label{sec:methods}
In this section, we give a brief explanation of persistent homology \cite{Edelsbrunner:2002,Zomorodian:2005} before introducing topological representations of protein-ligand binding interactions and protein   stability changes upon mutation.  
Multichannel topological deep learning and multitask topological deep learning architectures are constructed for binding affinity and mutation impact predictions. 

\subsection{Persistent homology}\label{sec:PH}

Simplicial homology gives a computable way to distinguish one space from another in topology and is built on   simplicial complex to extract topological invariants in a given data set. A simplicial complex $K$ is a topological space that is constructed from geometric components of a data set, including  discrete vertices (nodes or atoms in a protein), edges (line segments or bonds in a biomolecule), triangles,  tetrahedrons and their high dimensional counterparts, under certain rules.   Specifically, a 0-simplex is a vertex, a 1-simplex  an edge, a 2-simplex  a triangle, and a 3-simplex represents a tetrahedron.  The identification of connectivity of a given data set can follow different rules which leads to, for example,  Vietoris-Rips complex (VRC), C$\check{e}$ch complex and alpha complex. The linear combination of $k$-simplexes is called $k$-chain, which is introduced to associate the topological space,  i.e., simplicial complex, with algebra groups, which further facilitate the computation of the topological invariants (i.e., Betti numbers) in a given data set. Specifically, the set of all $k$-chains of simplicial complex $K$  is regarded as elements of a  chain group, an  abelian group, together with a modulo-2 addition operation rule. Loosely speaking, a boundary operator is defined to systematically eliminate one vertex from the $k$-simplex at a time, which leads to a family of abelian groups, including the $k$th cycle group and the $k$th boundary group. Then  the quotient group of the $k$th cycle group and the $k$th boundary group is called the $k$th homology group.  Then, 
the $k$th Betti number is computed the  rank of the $k$th homology group. 

Persistent homology is constructed via a filtration process, in which the connectivity of the given data set is systematically reset according to a scale parameter. More specifically, a  nested sequence of subcomplexes is defined via a filtration parameter, such as the growing radius of protein atoms located at their initial coordinates. For each  subcomplex, homology groups and corresponding Betti number can be computed. Therefore, the evolution of topological invariants over the filtration process can be recorded as  a barcode  \cite{Ghrist:2008} or a persistence diagram.   For a given data set, barcodes represent the persistence of its topological features over different spatial scales.   

\subsection{Topological representation of biomolecules}

\paragraph{Topological fingerprints}
A basic assumption of using persistent homology for biomolecular function prediction is that 1D biomolecular persistent barcodes are able to effectively characterize 3D biomolecular structures. We called such barcodes as topological fingerprints (TFs) \cite{KLXia:2014c, KLXia:2015a}.   
Fig.  \ref{fig:MutationBarcodes} illustrates the TFs of a wild type protein (PDB:1hmk) and its mutant  obtained from persistent homology calculations using the VR complex. The mutation (W60A) occurred at residue 60 from Trp to Ala is shown at Fig.  \ref{fig:MutationBarcodes}{\bf a} and {\bf b}. Apparently, a large residue (Trp) at the protein surface is replaced by a relatively small one (Ala). The corresponding barcodes are given  in Fig.  \ref{fig:MutationBarcodes} {\bf c} and {\bf d}, where three panels from top to bottom are for  Betti-0, Betti-1, and Betti-2, respectively. The barcodes for the wild type are  generated using heavy atoms within 6\AA~ from the mutation site. 
The mutant barcodes are obtained with the same set of heavy atoms in the protein except for those in the mutated residue.   
In two Betti-0 panels, their difference in the number of bars is equal to the difference in number of heavy atoms between the wild type and mutant.  
Broadly speaking,  the lengths of short bars reflect the  bond length of the corresponding heavy atom. Therefore, in both the wild type protein and the mutant, bond lengths for most heavy atoms are smaller than 1.8\AA. Additionally, bars that end between 1.8\AA ~ and 3.8 \AA~ might correlate with  hydrogen bonds.  By a comparison between  {\bf c} and {\bf d}, one can easily note the increase in the number of bars that end in the range of 1.8 - 3.8 \AA~ in the mutant, which might indicate a mutation induced steric effect.  In  Betti-1 and Betti-2 panels, the mutant has fewer bars than the wild type does because a smaller surface residue at 60 creates fewer ring and cavity contacts with the rest of the protein.  
 
The all heavy atom topological representation of proteins does not provide enough biological information about protein structures, such as bond length distribution of a given type of atoms, hydrogen bonds, hydrophobic and hydrophilic effects, etc. Therefore, we introduce element specific topological fingerprint (ESTF) to offer a more detailed characterization of protein-ligand binding and protein mutation. For example, Betti-1 and Betti-2 ESTFs from carbon atoms are associated with  hydrophobic interaction networks in biomolecules. Similarly ESTFs between  nitrogen and oxygen atoms correlate to hydrophilic interactions and/or hydrogen bonds in biomolcules. However,  hydrogen atoms are typically absent from structures in the PDB and thus are not used in our data driven ESTF description.  For proteins, commonly occurring heavy atom types include ${\rm C, N, O,}$ and ${\rm S}$. For ligands, we use 9 commonly occurring atom types, namely ${\rm C, N, O, S, P, F, Cl, Br,}$ and  ${\rm I}$. To characterize the interactions between protein and ligand binding, we construct cross protein-ligand ESTFs such that one type of heavy atoms is chosen from the protein and the other from the ligand. Therefore, there are a total of  thirty six sets of  ESTFs in each topological dimension. For mutation characterization, we describe the interactions between mutated residue and the rest of the protein and arrive at 16 sets of ESTFs in each topological dimension. Similarly, we 
we generate  16 sets of cross ESTFs in each topological dimension from the wild type protein to study the interactions between the residue to be mutated and the rest of the protein. To contrast the ESTFs of wild type protein and mutant, we take the differences between the above ESTFs, which gives rise to another 16 sets of ESTFs in each topological dimension. However, high dimensional Betti-1 and Betti-2 invariants require high the formation of high order complexes. As non-carbon atoms do not occur very often,  Betti-1 and Betti-2 ESTFs are omitted for non-carbon atoms, except specified.

Based on the above analysis of TFs and ESTFs, it is convenient to characterize them by $\mathbb{B}(\alpha,{\mathcal C},{\mathcal D})$ 
with $\alpha$ labeling the selection of atoms depending on atom types and affiliations (i.e., protein, ligand or mutated residue). Here ${\mathcal C}$ denotes the type of simplicial complex (i.e.,  Vietoris-Rips complex (VRC)  or alpha complex), and ${\mathcal D}$ indicates the dimension, such as Betti-0, Betti-1, or Betti-2.  Additionally, as shown in 	Fig.   \ref{fig:MutationBarcodes}, it is important to take a note on the birth, death, and persistence of each barcode, because this information is associated with the bond length, ring or cavity size, flexibility and steric effect. To this end, we use $\mathbf{V}^{\rm{b}}$, $\mathbf{V}^{\rm{d}}$, and $\mathbf{V}^{\rm{p}}$ to respectively represent birth, death, and persistence of barcodes. Moreover, Jeffrey argued that there are strong, moderate and weak hydrogen bond interactions with   donor-acceptor distances of 2.2-2.5\AA, 2.5-3.2\AA, and 3.2-4.0\AA, respectively  \cite{Jeffrey:1997}. Therefore, it is important to divide the filtration interval $[0,L]$   into $n$ equal length subintervals then characterize $\mathbf{V}^{\rm{b}}$, $\mathbf{V}^{\rm{d}}$, and $\mathbf{V}^{\rm{p}}$ accordingly.
\begin{equation}\label{eq:DigitalRepresentation}
\begin{aligned}
\mathbf{V}^{\rm{b}}_i &= \left\|\{(b_j,d_j)\in\mathbb{B}(\alpha,{\mathcal C},{\mathcal D})|(i-1)L/n\leq b_j\leq iL/n\}\right\|, \, 1\leq i < n, \\
\mathbf{V}^{\rm{d}}_i &= \left\|\{(b_j,d_j)\in\mathbb{B}(\alpha,{\mathcal C},{\mathcal D})|(i-1)L/n\leq d_j\leq iL/n\}\right\|, \, 1\leq i < n, \\
\mathbf{V}^{\rm{p}}_i &= \left\|\{(b_j,d_j)\in\mathbb{B}(\alpha,{\mathcal C},{\mathcal D})|(i-1)L/n\geq b_j,\,iL/n \leq d_j\}\right\|, \, 1\leq i \leq n,
\end{aligned}
\end{equation}
where $\|\cdot\|$ is cardinality of sets. Here $b_j,\, d_j$ are birth and death of bar $j$. The three types of representation vectors are computed for sets of Betti-1 and Betti-2 bars. For Betti-0 bars, since their births positions are uniformly $0$, only $\mathbf{V}^{\rm{d}}$ needs to be computed.  
To characterize pairwise interactions between atoms, it is convenient to simply use pairwise distance information between atoms. The corresponding image like representation denoted by $\mathbf{V}^{\rm{r}}$ can be constructed similarly to $\mathbf{V}^{\rm{d}}$ by substituting the set of barcodes by a collection of distances between the atom pairs of interest. It should be noted that $\mathbf{V}^{\rm{r}}$ is not equivalent to $\mathbf{V}^{\rm{d}}$ in most simplicial complex setups. Generally speaking, $\mathbf{V}^r$ also reflects the $0$th order topological connectivity information. It is used as the characterization of $0$th order connectivity of the biomolecules in the applications shown in this work. 
Finally, we denote $X_s$ all the feature vectors for the $s$th sample and $Y_s$ the corresponding target.

\paragraph{Image-like  multichannel topological representation}

To feed the outputs of TFs into convolutional neural network, the barcodes are transformed to a 1D image like representation with multiple channels. Topological feature vectors , $\mathbf{V}^{\rm{b}}$, $\mathbf{V}^{\rm{d}}$, and $\mathbf{V}^{\rm{p}}$, can be viewed as a one-dimensional (1D) image. Each each subinterval in the filtration axis represents a digit (or pixel) in the 1D image like representation. Such a treatment of topological features describes the topological information with appropriately chosen resolution of $L/n$.  Meanwhile, the chemical information in the ESTFs of $\mathbb{B}(\alpha,{\mathcal C},{\mathcal D})$ are described by multiple channels in the 1D image representation, which is similar to the  RGB color image representation. However, in our description, each pixel is associated with $m$ channels to describe different element type, protein mutation status (i.e., wild type and mutant), topological dimension (i.e., Betti-0, Betti-1 and Betti-2), and topological event  (i.e.,  birth, death, and persistence). Each element in the 1D image like representation is standardized to have zero mean and unit variance among the data sets. This 1D image-like topological representation can be easily transferred among problems such as protein-ligand binding affinity modeling and prediction of protein stability change upon mutation while traditional machine learning approach requires manual extraction of features for each domain of application. When convolutional neural network is applied, the convolution layers identify local patterns of atomic interactions and the fully connected layers then extract higher order descriptions of the system by combining local patterns at various distance scales.


\subsection{Multichannel topological convolutional neural network (MT-TCNN)}

\begin{figure}[ht]
\begin{center}
\includegraphics[keepaspectratio,width=5in]{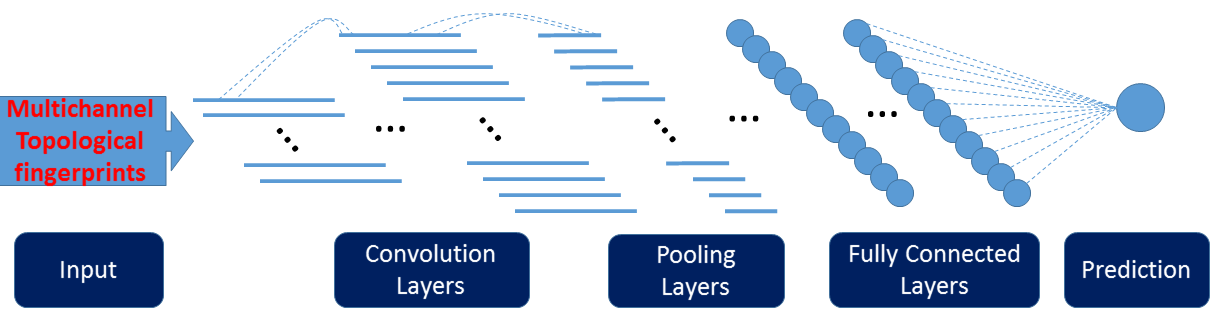}
\caption{An illustration of the 1D convolutional neural network used with repeated convolution layers and pooling layers followed by several fully connected layers.}
\label{sup:Figure_PHDLBP_net}
\end{center}
\end{figure}

The preprocessed multichannel topological image is standardized with mean 0 and standard deviation 1 for being used in convolutional neural networks. A convolutional neural network with a few 1D convolution layers followed by several fully connected layers is used to extract higher level features from multichannel topological images  and to perform regression with the learned features. An illustration of the convolutional neural network structure is shown in Fig. \ref{sup:Figure_PHDLBP_net}. A brief review of  multichannel topological convolutional neural network concepts is given in the case of 1D image like TFs. Convolution operation, optimization method for feedforward neural networks, and dropout out technique which prevents overfitting are discussed. One of the advantages of  multichannel  topological convolutional deep neural networks is their capability of extracting features hierarchically from low level topological representations. 

\paragraph{Convolution operation}

Consider an $n\times m$ second order tensor $\mathbf{V}$, where $n$ is the number of  topological feature pixels and $m$ is number of channels for each pixel. In this approach, $n$ corresponds to the radius filtration dimension of the biomolecule topological analysis and $m$ corresponds the number of representation vectors used which are defined in Eq. (\ref{eq:DigitalRepresentation}). With a predefined window size $w$, a convolutional filter $\mathbf{F}$ can be represented by a $w\times m$ second order tensor. By moving the window of size $w$ along the  radius filtration direction of $\mathbf{V}$, a sequence of $N_f$ second order tensors which are subtensors of $V$ are obtained and can be concatenated to form an $N_f\times w\times m$ third order tensor $\mathbf{T}$. The filter $\mathbf{F}$ operated on $\mathbf{T}$ results in a first order tensor $\mathbf{T}_{ijk}\mathbf{F}_{jk}$ by tensor contraction. Concatenating the outputs of $n_f$ filters gives an $N_f\times n_f$ second order tensor. Generally speaking, a 1D convolution layer takes in an $n\times m$ tensor and outputs an $N_f\times n_f$ tensor. 

\paragraph{Optimization}
Feedforward neural networks are usually trained by backpropagation where the error of the output layer is calculated and is propagated backward through the network to update its weights. For structured neural networks, conventional $L_2$ minimization does not work.  
One popular approach of training a neural network is the stochastic gradient decent (SGD) method. Let $\Theta$ be the parameters in the network and ${\mathcal L}(\Theta)$ be the objective function or learning kernel that is to be minimized. SGD method updates $\Theta_i$ to $\Theta_{i+1}$ from step $i$ to step $i+1$ as
\begin{equation}\label{eq:SGD}
\Theta_{i+1}=\Theta_i- \tau\nabla_\Theta {\mathcal L}(\Theta_i; X_s,Y_s),
\end{equation}
where $\tau$ is the learning rate, $X_s$ and $Y_s$ are the input and target of the $s$th sample of the training set. In practice, the training set $(X,Y)$ is often split into mini-batches $\{(X_s,Y_s)\}_{s\in S}$. SGD method then goes through each mini-batch at a time instead of going through only one example at a time. When the landscape of the objective function is like a long steep valley, momentum is added to accelerate convergence of the algorithm. We therefore change the updating scheme to 
\begin{equation}\label{eq:SGDMomentum}
\begin{aligned}
\Delta\Theta_{i} &= \Theta_{i}-\Theta_{i-1}, \\
\Theta_{i+1} &= \Theta_i-(1-\eta)\tau\nabla_\Theta {\mathcal L}(\Theta_i; X^i_s,Y^i_s)+\eta\Delta\Theta_{i},
\end{aligned} 
\end{equation}
where $0\leq \eta \leq 1$ is a scalar coefficient for the momentum term.

\paragraph{Dropout} 
Neural networks with several convolution layers and fully connected layers possesses a large number of degrees of freedom which can easily lead to overfitting. Dropout technique is an easy way of preventing network overfitting \cite{srivastava2014dropout}. During training process, the hidden units are randomly chosen to feed zero values to their connected neighbors in the next layer. Suppose that a percentage   of neurons at a certain layer are chosen to be dropped during training, in the testing process, the output of this layer is computed by multiplying a coefficient such as $1-\lambda$, where $ \lambda$ is the dropout rate,  to approximate the average of the network after dropout in each training step.

\paragraph{Bagging (bootstrap aggregating)}
In addition to dropout technique which regularizes each individual model, bagging is a technique to combine the output of several models trained separately by averaging to reduce generalization error based on the assumption that models with randomness in the training process likely make different errors on testing data. Generally, bagging method trains different models on different subsets of the training set. Specifically, as neural networks have relatively high underlying randomness caused by factors including the random weights initialization and the random mini-batch partition, it can benefit from bagging even if the individual models are trained on the same dataset. In this work, bagging of neural network models trained individually with the same architecture and training dataset is used.


\paragraph{Multitask   deep learning}

\begin{figure}[ht]
\begin{center}
\includegraphics[keepaspectratio,width=5.0in]{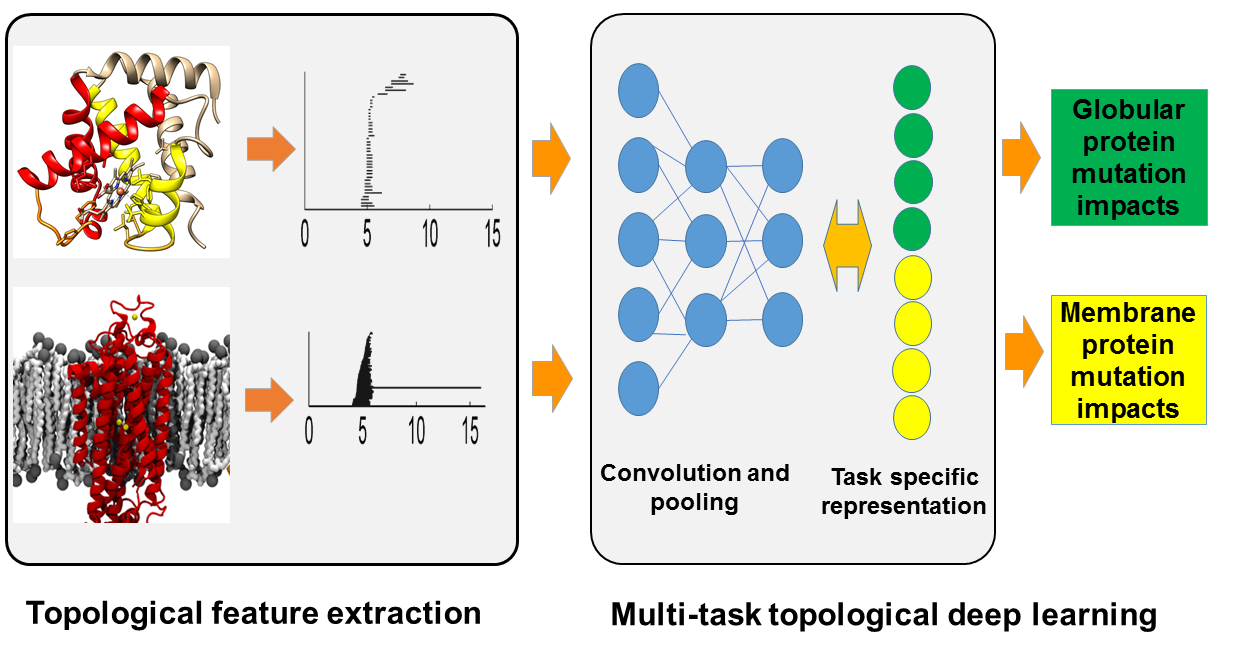}
\end{center}
\caption{Illustration of  multitask topological deep learning scheme
that shares and transforms topological information for the  simultaneous training and prediction of  
 globular protein and membrane protein mutation impacts. 
}\label{fig:deep_mutlitask}
\end{figure}

We construct a  multitask and multichannel topological deep learning  architecture to carry out simultaneous predictions. The common topological  attributes and underlying physical interactions in features provide a basis  for multitask predictions. Because the deep neoral networks are jointly trained from multiple prediction tasks, we expect the networks to generate robust high-level representations from low level TFs for prediction problems. We also expect that the refined representation would lead to prediction models with improved generalized performance. From the proposed deep learning models, we hope to  gain insights into how the nonlinearly and nonlocal  interactions among topological features  impact  various prediction tasks, which could further lead to better understanding towards the interactions among biomolecular prediction tasks. 
Finally, tasks with insufficient training data sets will be more likely to benefit from the information collected from tasks with large training sets in a multitask learning framework. Fig. \ref{fig:deep_mutlitask} illustrates our multitask topological deep learning  architecture for   simultaneous training and prediction of   globular protein and membrane protein mutation impacts. Convolutional deep neural networks are used.  

 The objective function for multitask minimization can be decomposed into training loss, similarity penalty for shared layers, and regularization term as 
\begin{equation}\label{ObjectiveFunction}
\begin{aligned}
{\mathcal L}(\Theta;X,Y) &= \sum_{j=1}^N{\mathcal J}_j (\Theta_{Sj},\Theta_{Bj};X_{j},Y_{j}) \\
&+ {\mathcal P}(\Theta_{S1},\cdots, \Theta_{SN}) \\
&+ {\mathcal R}(\Theta),
\end{aligned}
\end{equation}
where $\Theta$ is the collection of all parameters to be updated, $\Theta_{Sj}$ is the set of parameters for the $j$th task of the shared layers, $\Theta_{Bj}$ is the set of   parameters for the $j$th branch of neurons dedicated for the $j$th task,    and $(X_{j}, Y_j)$ are training data for the $j$th task. Here  ${\mathcal P}$ is the penalty function which penalizes the difference among $N$ sets of parameters. Finally ${\mathcal R}(\cdot)$ is the regularization term which prevents overfitting and ${\mathcal J}$ is the $j$th loss function.

\section{Results}\label{sec:results}

\subsection{Deep learning prediction of protein-ligand binding affinities}

Protein-ligand binding is a fundamental  biological process in cells and involves detailed molecular recognition, synergistic protein-ligand interaction, and possibly protein conformational changes.  Agonist binding is crucial to receptor  functions and typically triggers a physiological response, such as transmitter-mediated signal transduction, hormone and growth factor regulated metabolic pathways,  stimulus-initiated gene expression, enzyme production, cell secretion, etc.  The understanding of protein-ligand interactions has been a fundamental issue in molecular biophysics, structural biology and medicine. A specific issue in drug and protein design is to predict protein-ligand binding affinity from given structural information   \cite{Kuntz:1982,DesJarlais:1986,Goodsell:1990,Jorgensen:1991, Kollman:2000ACR,gilson1997statistical, Gilson:2007}.
Protein-ligand binding affinity is a measurement of rate of binding which indicates the degree of occupancy of a ligand at the corresponding protein binding site and is affected by several factors including intermolecular interaction strength and solvation effects. The ability of predicting protein-ligand binding affinity to a desired accuracy is a prerequisite for the success of many applications in biochemistry such as protein-ligand docking and drug discovery. 
Broadly speaking, there are three types of binding affinity predictors (commonly called scoring functions): physics based \cite{Ortiz:1995,Yin:2008}, empirical \cite{Zheng:2015LISA,Verkhivker:1995PLP, Eldridge:1997,WangRenXiao:2002,Zheng:2013MoveableType,PMFScore:1999, DrugScore:2005,ITScore:2006}, and knowledge based  \cite{HongjianLi:2014RF,Sarah:2011,Ashtawy:2012}.  In general, physics based scoring functions invoke QM and QM/MM approaches \cite{CHARMM22, AMBER15} to provide unique insights into the molecular mechanism of protein-ligand interactions. 
A prevalent view is that binding involves intermolecular forces, such as steric contacts, ionic bonds, hydrogen bonds, hydrophobic effects and van der Waals interactions. 
Empirical scoring functions work well but require carefully selected data sets and parametrization \cite{Zheng:2015LISA,Verkhivker:1995PLP, Eldridge:1997,WangRenXiao:2002}. However, both  physics based scoring functions and empirical scoring functions employ linear superposition principles that are  not explicitly designed to deal with exponentially  growing and increasingly diverse  experimental data sets. 
Knowledge based scoring functions use  modern machine learning techniques, which utilize nonlinear regression and exploit large data sets to uncover  underlying patterns within the data sets. 
Given the current massive and complex data challenges, knowledge based scoring functions outperform all other scoring functions
 \cite{Zheng:2015LISA}.

\begin{figure}[ht]
\begin{center}
\includegraphics[keepaspectratio,width=4in]{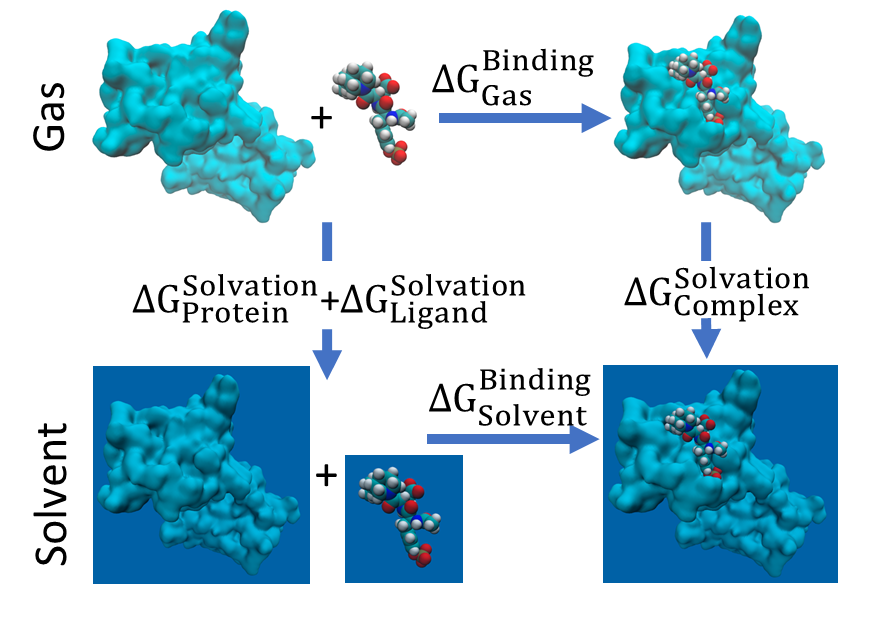}
\caption{Energy cycle of protein-ligand binding free energy modeling.}
\label{fig:BindingEnergy}
\end{center}
\end{figure}

 In computation, the binding affinity or alternatively the binding free energy can be modeled via an energy cycle as shown in Fig. \ref{fig:BindingEnergy} where the main contributors to the process are intermolecular interactions and solvation effects. In this work, we consider the set of element types $\mathbb{L}^{\rm{e}}=\{\rm C,N,O,S,P,F,Cl,Br,I\}$ in ligands and $\mathbb{P}^{\rm{e}}=\{\rm C,N,O,S\}$ in proteins. We define an opposition distance between two atoms $a_i$ and $a_j$ as 
\begin{equation}\label{eq:Distance}
d^{op}(a_i,a_j)=
\begin{cases}
d(a_i,a_j)&, A(a_i)\neq A(a_j)\\
\infty&, A(a_i)=A(a_j)
\end{cases},
\end{equation}
where $d(\cdot,\cdot)$ is Euclidean distance between two atoms and $A(\cdot)$ denotes the affiliation of an atom which is either a protein or a ligand. 

The ESTFs used in this application are summarized in Table \ref{tab:BindingRepresentation}. ESTFs are generated according to the definition given in Eq. (\ref{eq:DigitalRepresentation}). As shown in Table \ref{tab:BindingRepresentation}, five sets of ESTFs are constructed. The differences between Set 2 and Set 3 as well as Set 4 and Set 5  are also employed as representation vectors to address the impact of ligand binding resulting in a total of 72 representation vectors (i.e., channels) forming the 1D image like representation of the protein-ligand complex. Here, 0-dimensional TFs describe  intramolecular interactions between the protein and ligand.  All heavy atom TFs delineate the geometric effect of protein-ligand binding.    The TFs of carbon atoms account for hydrophobic effects and also implicitly reflect the solvation effects. 

\begin{table}[ht]
\caption{List of topological representations of protein-ligand complexes.  $\mathbb{P}$ and $\mathbb{L}$ are sets of atoms in protein and in ligand. $T(\cdot)$ denotes element type of an atom. $e_{\rm{P}}$ is an element type in protein and $e_{\rm{L}}$ is an element type in ligand.   
``Complex'' refers to the type of simplicial complex used and ``Dimension''  refers to the dimensionality of a topological invariant.}
\label{tab:BindingRepresentation}
\centering
\begin{tabular}{|l|l|l|l|l|}
\hline
 Set & Atoms used & Distance & Complex & Dimension \\
\hline
1 & $\{a\in \mathbb{P}|T(a)=e_{\rm{P}}\}\cup\{a\in \mathbb{L}|T(a)=e_{\rm{L}}\}, e_{\rm{P}}\in\mathbb{P}^{\rm{e}}, e_{\rm{L}}\in\mathbb{L}^{\rm{e}}$& $d^{op}$ & - & 0 \\
\hline
2 & $\{a\in\mathbb{P}|T(a)\in\mathbb{P}^{\rm{e}}\}$ & Euclidean & Alpha & 1,2 \\
\hline
3 & $\{a\in\mathbb{P}|T(a)\in\mathbb{P}^{\rm{e}}\}\cup\{a\in\mathbb{L}|T(a)\in\mathbb{L}^{\rm{e}}\}$ & Euclidean & Alpha & 1,2 \\
\hline
4 & $\{a\in\mathbb{P}|T(a)=\rm{C}\}$ & Euclidean & Alpha & 1,2 \\
\hline
5 & $\{a\in\mathbb{P}|T(a)=\rm{C}\}\cup\{a\in\mathbb{L}|T(a)=\rm{C}\}$ & Euclidean & Alpha & 1,2 \\
\hline
\end{tabular}
\end{table}

Due to the huge amount of computation resources required, we repeatedly train 100 single neural networks individually. To test the performance of bagging of the models, we randomly select 50 trained models from the 100 and output the performance for the averaged predictions. This process is repeated 100 times and both median and best results are reported.

In this example, the proposed method is tested on the PDBBind 2007 data set \cite{PDBBind:2015}. The  PDBBind 2007 core set of 195 protein-ligand complexes  is used as the test set and  the PDBBind 2007 refined set, excluding the  PDBBind 2007 core set, is used as the training set with 1105 protein-ligand complexes. A comparison between the proposed method and other binding affinity predictors is summarized in Table \ref{tab:BindingPerformance}. Clearly, the present topology based network binding predictor (TNet-BP) outperforms all the other scoring functions reported by  Li \emph{et al} \cite{HLi:2015}.  

\begin{table}[ht]
\centering
\rowcolors{2}{gray!25}{white}
\begin{tabular}{lll}
\toprule
\rowcolor{gray!75}
Method & $R_P$ & RMSE \\ 
\midrule
TNet-BP & \BindPCCMedian$^a$ & \BindRMSEMedian \\
RF::VinaElem & 0.803 & 1.42 \\
RF:Vina & 0.739 &1.61 \\ 
Cyscore & 0.660 & 1.79 \\ 
X-Score::HMScore & 0.644 & 1.83 \\
MLR::Vina & 0.622 & 1.87 \\
HYDE2.0::HbondsHydrophobic & 0.620 & 1.89 \\
DrugScore & 0.569 & 1.96 \\
SYBYL::ChemScore & 0.555 & 1.98 \\
AutoDock Vina & 0.554 & 1.99 \\
DS::PLP1 & 0.545 & 2.00 \\
GOLD::ASP & 0.534 & 2.02 \\
SYBYL::G-Score & 0.492 & 2.08 \\
DS::LUDI3 & 0.487 & 2.09 \\
DS:LigScore2 & 0.464 & 2.12 \\
GlideScore-XP & 0.457 & 2.14 \\
DS::PMF & 0.445 & 2.14 \\
GOLD::ChemScore & 0.441 & 2.15 \\
PHOENIX & 0.616 & 2.16 \\
SYBYL::D-Score & 0.392 & 2.19 \\
DS::Jain & 0.316 & 2.24 \\
IMP::RankScore & 0.322 & 2.25 \\
GOLD::GoldScore & 0.295 & 2.29 \\
SYBYL::PMF-Score & 0.268 & 2.29 \\
SYBYL::F-Score & 0.216 & 2.35 \\
\bottomrule
\end{tabular}
\caption{Comparison of optimal Pearson's correlation coefficients $R_P$ and  RMSEs ($pK_d/pK_i$) of various scoring functions for the prediction of protein-ligand binding affinity of the PDBBind 2007 core set. Except for the result of our TNet-BP, all other results are adopted from Li \emph{et al} \cite{HLi:2015}. Best performance is reported for empirical scoring functions with optimal selection of parameters. $^a$ Median results (The best $R_P=\BindPCCBest$ and best RMSE=$\BindRMSEBest$ for this method).
}
\label{tab:BindingPerformance}
\end{table}

\subsection{Deep learning prediction of protein folding free energy changes upon mutation}

Aside from some unusual exceptions \cite{Schroder:2005,Chiti:2006}, proteins fold in particular three-dimensional structures to provide the structural basis for living organisms \cite{Anfinsen:1973}. Protein functions, i.e., acting as enzymes, cell signaling mediators, ligand receptors and structural supports, are typical   consequences of a delicate balance between protein structural stability and flexibility.  
Mutation that changes protein amino acid sequences through non-synonymous single nucleotide substitutions (nsSNPs) plays a fundamental role in selective evolution. Such substitutions may   lead to the loss or the modification of certain functions \cite{PYue:2005}. 
Mutations are often associated with various human diseases \cite{ZZhang:2012, Kucukkal:2015}, and 
For example, mutations in proteases and their natural inhibitors result in more than 60 human hereditary diseases \cite{Puente:2003}. Additionally, mutation is a general cause for  drug resistance  \cite{Martinez:2000}. 
Artificially designed mutations are used to understand mutation impacts to protein   structural stability, flexibility and function, as well as  mutagenic diseases, and evolution pathways of organisms   \cite{Fersht:2078}. However,  mutagenesis experiments  are typically costly and time-consuming. Computational prediction of mutation  impacts is able to systematically explore protein structural instabilities, functions, disease connections, and organismal evolution pathways \cite{Guerois:2002} and provide an economical, fast, and potentially accurate alternative to mutagenesis experiments.  Many  computational methods have been developed in the past decade,  including support vector machine  \cite{Capriotti:2005}, statistical potentials \cite{Dehouck:2009}, knowledge-modified MM/PBSA approach  \cite{Getov:2016},  (Rosetta-high) protocols \cite{Kellogg:2011},    FoldX (3.0, beta 6.1) \cite{Guerois:2002},   SDM \cite{Worth:2011},  DUET \cite{Pires:2014b},  PPSC (Prediction of Protein Stability, version 1.0) with the 8 (M8) and 47 (M47) feature sets \cite{YYang:2013},  PROVEAN \cite{YChoi:2012}, ELASPIC  \cite{Berliner:2014}, and EASE-MM \cite{Folkman:2016}.

\begin{figure}[ht]
\begin{center}
\includegraphics[keepaspectratio,width=4in]{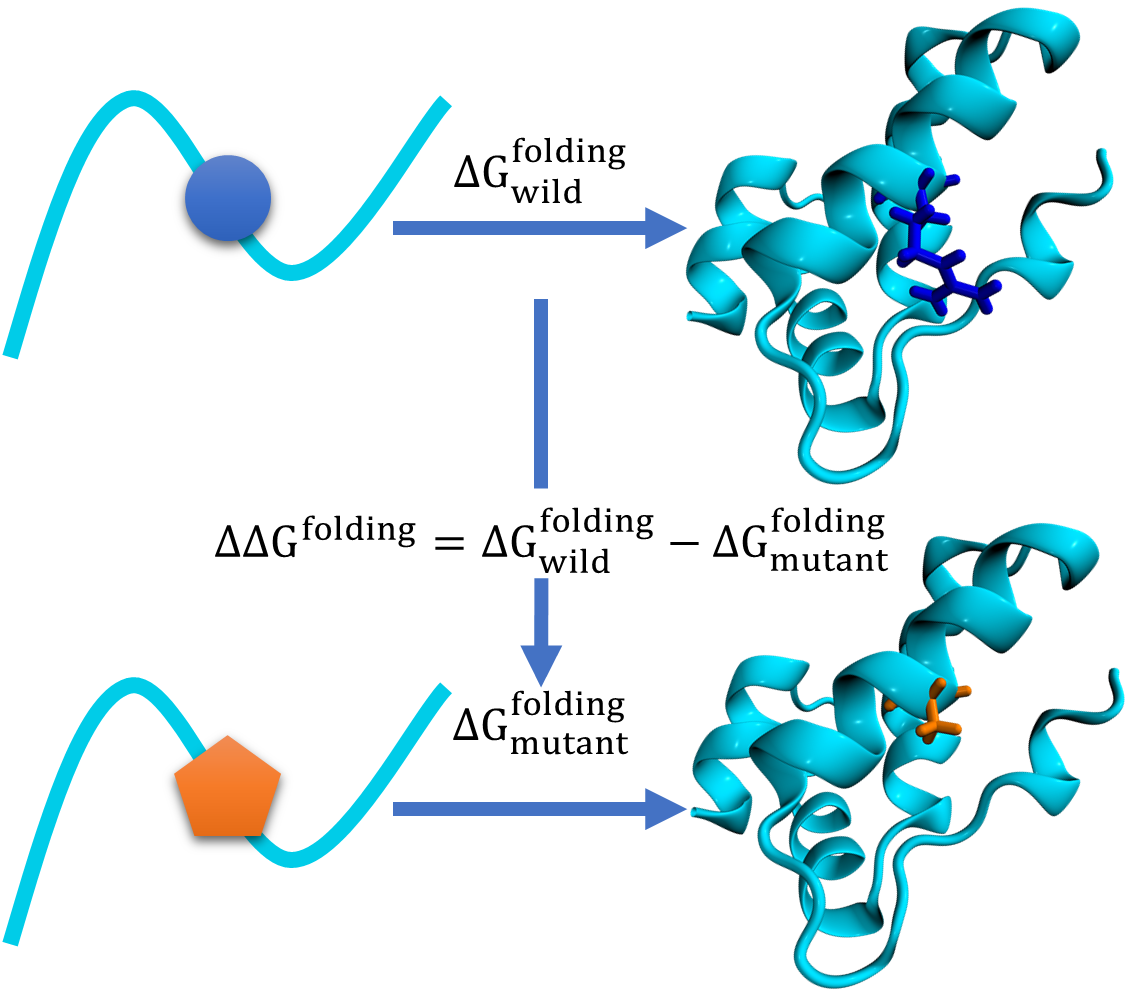}
\caption{Mutation induced   change in protein folding free energy.}
\label{fig:MutationEnergy}
\end{center}
\end{figure}
\begin{figure}
\begin{center}
\includegraphics[keepaspectratio,width=4.0in]{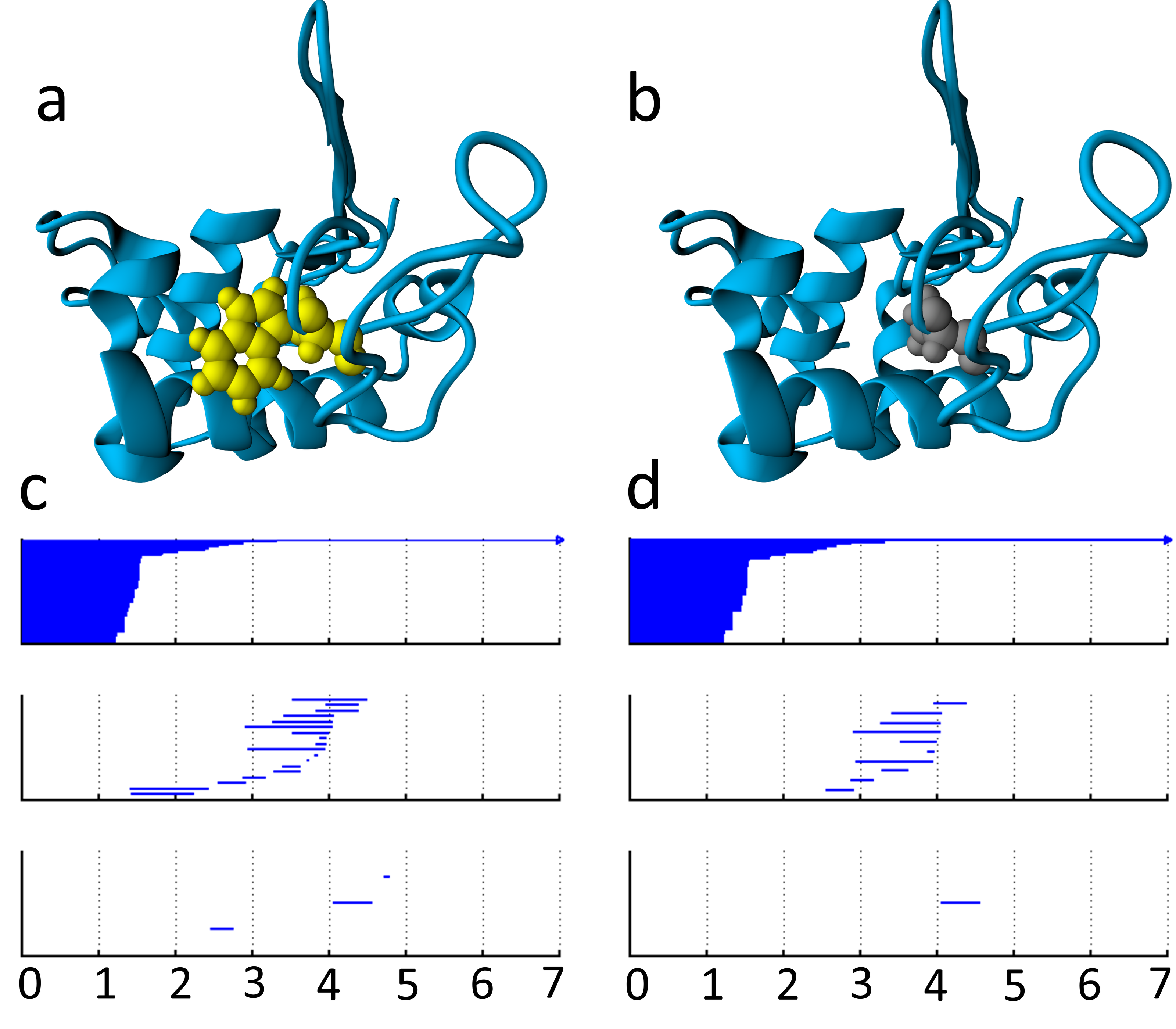} 
\caption{An illustration of   barcode changes from wild type  to  mutant proteins. 
\textbf{a} The wild type protein (PDB:1hmk) with residue 60 as Trp. 
\textbf{b} The mutant with residue 60 as Ala. 
\textbf{c} Wild type protein barcodes for  heavy atoms  within 6 \AA~ of the mutation site.   Three panels  from top to bottom are Betti-0, Betti-1, and Betti-2 barcodes, respectively. The horizontal axis is the filtration radius (\AA).  
\textbf{d} Mutant protein barcodes obtained similarly as those for  the wild type.
}
\label{fig:MutationBarcodes}
\end{center}
\end{figure}

Modeling protein folding free energy change upon mutation basically involves the unfolded states and folded structures of mutant and wild type as shown in Fig. \ref{fig:MutationEnergy}. Since unfolded states of proteins are highly dynamical which significantly increases the modeling cost due to the need of sampling over large conformation space, we only analyze the folded states of mutants and wild type proteins in this application. Similar to the protein-ligand binding affinity prediction, atomic interactions between specific element types, geometric effects, and hydrophobic effects are characterized. The persistent homology analysis performed in this application is summarized in Table \ref{tab:MutationRepresentation}. The differences between Sets 1 and 2 as well as 3 and 4 are also included to account for changes caused by mutation. The 1D image like representation in this application thus has a channel size of 45. An example of the persistent homology barcodes of a mutant and its wild type is given in Fig.  \ref{fig:MutationBarcodes}.

\begin{table}[ht]
\centering
\caption{List of topological representations of protein mutations. Here $\mathbb{P}^{\rm{W}}$, $\mathbb{P}^{\rm{M}}$, $\mathbb{M}^{\rm{W}}$, and $\mathbb{M}^{\rm{M}}$ are sets of atoms of wild type protein, mutant protein, mutation site in the wild type protein, and mutated site in the mutant protein. Here $\mathbb{P}^{\rm{e}}=\{\rm C, N, O \}$ and $T(\cdot)$ is the same as defined in Table \ref{tab:BindingRepresentation}. The distance function $d^{op}$ is similar to the one defined in Eq. (\ref{eq:Distance}), while the affiliation function $A(\cdot)$ returns either $\mathbb{M}$ or $\mathbb{P}\backslash\mathbb{M}$.}
\label{tab:MutationRepresentation}
\begin{tabular}{|l|l|l|l|l|}
\hline
 Set & Atoms selected & Distance & Complex & Dimension \\
\hline
1 & $\{a\in \mathbb{P}^{\rm{W}}\backslash\mathbb{M}^{\rm{W}}|T(a)=e_{\rm{P}}\}\cup\{a\in \mathbb{M}^{\rm{W}}|T(a)=e_{\rm{M}}\}, e_{\rm{P}},e_{\rm{M}}\in\mathbb{P}^{\rm{e}}$& $d^{op}$ & - & 0 \\
\hline
2 & $\{a\in \mathbb{P}^{\rm{M}}\backslash\mathbb{M}^{\rm{M}}|T(a)=e_{\rm{P}}\}\cup\{a\in \mathbb{M}^{\rm{M}}|T(a)=e_{\rm{M}}\}, e_{\rm{P}},e_{\rm{M}}\in\mathbb{P}^{\rm{e}}$& $d^{op}$ & - & 0 \\
\hline
3 & $\{a\in\mathbb{P}^{\rm{W}}|T(a)\in\mathbb{P}^{\rm{e}}\}$ & Euclidean & Alpha & 1,2 \\
\hline
4 & $\{a\in\mathbb{P}^{\rm{M}}|T(a)\in\mathbb{P}^{\rm{e}}\}$ & Euclidean & Alpha & 1,2 \\
\hline
\end{tabular}
\end{table}

Similar to the procedure used in the protein-ligand binding example, we repeatedly train 100 single neural networks individually for the tasks with defined train/test split. To test the performance of bagging of the models, we randomly select 50 trained models from the 100 and output the performance for the averaged predictions. This process is repeated 100 times and the median is reported. In the case of cross validation, 10 sets of 5-fold splits are generated randomly and 20 single models are generated for each split. The average prediction is taken over the 20 models within each split and the median result of the 10 splits are reported.

\begin{figure}[ht]
\begin{center}
\includegraphics[keepaspectratio,width=0.3\textwidth]{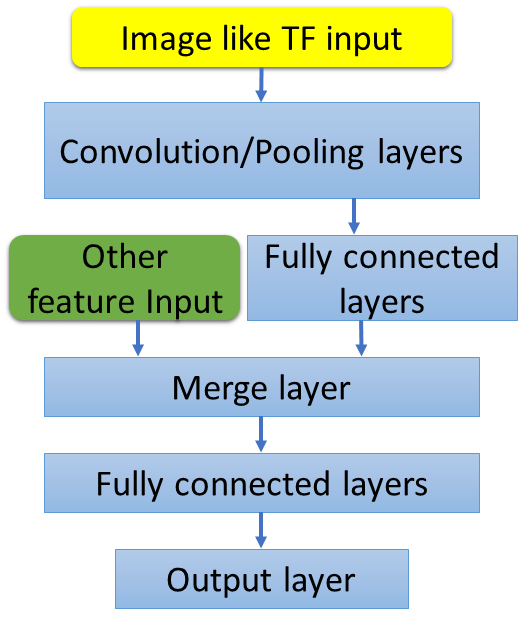}
\caption{An illustration of a deep learning architecture for incorporating non-image features in multichannel topological convolutional deep neural networks.}
\label{fig:Multifeature}
\end{center}
\end{figure}

The proposed method is tested on a data set of 2648 mutation instances of 131 proteins named ``S2648" data set \cite{Dehouck:2009} in a cross validation task over the ``S2648" set and a task of prediction of the ``S350" set which is a subset of ``S2648" set with the rest of the data as the training set. All thermodynamic data are obtained from the ProTherm database \cite{Bava:2004}. A comparison of the performance of various  methods is summarized in Table \ref{tab:MutationPerformance}.  Among them,  STRUM \cite{LJQuan:2016} is based on structural, evolutionary and sequence information and thus gives rise to excellent performance. We therefore have constructed two topology based neural network  mutation predictors (TNet-MPs).    TNet-MP-1 is solely based on topological information while   TNet-MP-2 has added with evolutionary and sequence  information, which is merged into the fully connected layer of the convolutional deep neural network, see Fig. \ref{fig:Multifeature}. TNet-MP-2 is able to significantly improve our original topological prediction, indicating the importance of  evolutionary and sequence information to mutation prediction. The details of the handcrafted features can be found in \cite{Tmutation}.

\begin{table}[ht]
\centering
\rowcolors{2}{gray!25}{white}
\begin{tabular}{lcccccc}
\toprule
\rowcolor{gray!75}
Method & \multicolumn{3}{c}{S350} & \multicolumn{3}{c}{S2648} \\
\rowcolor{gray!75}
 & $n^d$ & $R_P$ & RMSE  & $n^d$ & $R_P^e$ & RMSE$^f$ \\
\midrule
 TNet-MP-2 & 350 & \MutTestSeqPCCMedian & \MutTestSeqRMSEMedian & 2648 & \MutCVSeqPCCMedian & \MutCVSeqRMSEMedian\\
 STRUM$^b$        & 350 & 0.79 & 0.98 & 2647 & 0.77 & 0.94 \\
  TNet-MP-1 & 350 & \MutTestPCCMedian & \MutTestRMSEMedian & 2648 & \MutCVPCCMedian & \MutCVRMSEMedian\\
 mCSM$^{b,c}$     & 350 & 0.73 & 1.08 & 2643 & 0.69 & 1.07 \\
 INPS$^{b,c}$     & 350 & 0.68 & 1.25 & 2648 & 0.56 & 1.26 \\
 PoPMuSiC 2.0$^b$ & 350 & 0.67 & 1.16 & 2647 & 0.61 & 1.17 \\
 PoPMuSiC 1.0$^a$ & 350 & 0.62 & 1.23 & - & - & - \\
 I-Mutant 3.0$^b$ & 338 & 0.53 & 1.35 & 2636 & 0.60 & 1.19 \\
 Dmutant$^a$      & 350 & 0.48 & 1.38 & - & - & - \\ 
 Automute$^a$     & 315 & 0.46 & 1.42 & - & - & - \\
 CUPSAT$^a$       & 346 & 0.37 & 1.46 & - & - & - \\
 Eris$^a$         & 334 & 0.35 & 1.49 & - & - &  - \\
 I-Mutant 2.0$^a$ & 346 & 0.29 & 1.50 & - & - & - \\
\bottomrule
\end{tabular}
\caption{Comparison of  Pearson's correlation coefficients ($R_P$) and RMSEs (kcal/mol) of various methods on the prediction task of  the ``S350" set and 5-fold cross validation of the ``S2648". TNet-MP-1 is our topological based convolutional neural network model that solely utilizes topological information. TNet-MP-2 is our model that complements TNet-MP-1 with  manually extracted evolutionary and sequence information. $^a$ Data directly obtained from Worth \emph{et al}\cite{Worth:2011}. $^b$ Data obtained from Quan \emph{et al} \cite{LJQuan:2016}. $^c$ The results reported in the publications are listed in the table, however, according to Ref. \cite{LJQuan:2016},  the data from the online server has $R_p$ (RMSE) of 0.59 (1.28) and 0.70 (1.13) for INPS and mCSM respectively in the task of S350 set. $^d$ Number of samples successfully processed.  }
\label{tab:MutationPerformance}
\end{table}

\subsection{Multitask deep learning prediction of membrane protein mutation impacts}

\begin{figure}[ht]
\begin{center}
\includegraphics[keepaspectratio,width=0.4\textwidth]{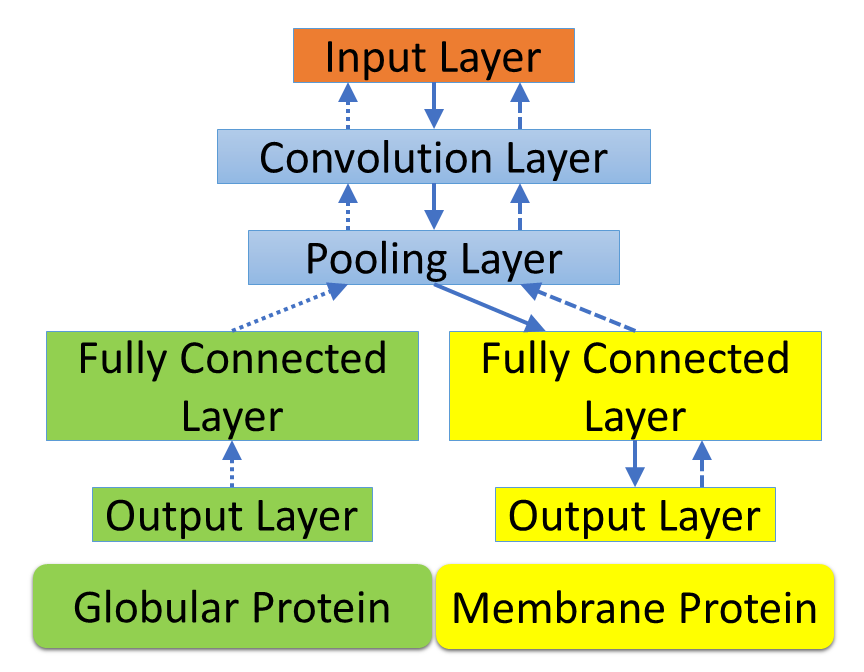}
\caption{An illustration of a multitask deep learning architecture for  using globular protein stability change upon mutation as an auxiliary task to improve the task of membrane protein mutation prediction. The solid arrows show the path of information passing when the model is applied for predictions. The dotted and dashed arrows mark the paths of backpropagation when the network is trained with globular protein data set and membrane protein data set respectively.}
\label{fig:MTLMembrane}
\end{center}
\end{figure}

Multitask learning offers an efficient way to improve the predictions having small data size by  taking the advantage of larger data sets 
 \cite{zhou2011malsar}.   Although a large amount of thermodynamic data is available for globular  protein mutations, the mutation data set for membrane proteins is relatively small, which stands between 200 and 300 \cite{Kroncke:2016}. The small size of membrane protein mutation data limits the success of data driven approaches, such as ensemble of trees.  While the popular multitask learning framework built on linear regression with regularization techniques lacks the ability of extracting relationship between very low level descriptors and  target quantity, neural network with a hierarchical structure provides a promising option for such problems. We add the prediction of protein stability change upon mutation for globular proteins as an auxiliary task for the task of prediction of membrane protein stability changes upon mutation. In the designed network architecture shown in Fig. \ref{fig:MTLMembrane},  two tasks share convolution layers and the network splits into two branches with fully connected layers for two tasks. Intuitively, the task of globular protein mutation predictions  helps  the extraction of higher level features from low level topological representations and the branch for membrane protein mutation predictions learns the feature-target relationship from the learned high level features. 
 
The proposed method is tested on a set of 223 mutation instances of membrane proteins covering 7 protein families named ``M223" data set   \cite{Kroncke:2016} with 5-fold cross validation. A comparison with other methods is shown in Table \ref{tab:MemMutationPerformance}. 
Pearson's correlation coefficient of membrane protein mutation prediction is improved 8.3\%, i.e., from 0.48 to 0.52. 
As noted by Kroncke {\it et al}, there is no reliable methods for the prediction of  membrane protein mutation impacts at the present \cite{Kroncke:2016}. Our TopologyNet results, though are not not satisfactory, are the best for this problem. 

\begin{table}[ht]
\centering
\rowcolors{2}{gray!25}{white}
\begin{tabular}{lll}
\toprule
\rowcolor{gray!75}
Method & $R_P$ & RMSE\\
\midrule
TNet-MMP-2$^d$ & \MemMTLPCCMedian & \MemMTLRMSEMedian\\
TNet-MMP-1$^c$ & \MemSinglePCCMedian & \MemSingleRMSEMedian\\
Rosetta-MP         & 0.31 & - \\
Rosetta (High)$^a$ & 0.28 & - \\
FoldX              & 0.26 & 2.56 \\
PROVEAN            & 0.26 & 4.23 \\
Rosetta-MPddG      & 0.19 & - \\
Rosetta (low)$^b$  & 0.18 & - \\
SDM                & 0.09 & 2.40 \\
\bottomrule
\end{tabular}
\caption{Comparison of  Pearson's correlation coefficients ($R_P$) and RMSEs (kcal/mol) of various methods for the ``M223" data set. Except for the present results for TNet-MMP-1 and TNet-MMP-2, all other results are adopted from Kroncke \emph{et al}\cite{Kroncke:2016}. The results of Rosetta methods are obtained from Fig. S1 of Ref. \cite{Kroncke:2016} where RMSE is not given. The results of other methods are obtained from Table S1 of Ref. \cite{Kroncke:2016}. The results of the machine learning based methods are not listed since those servers are not trained on membrane protein data sets. Among the methods listed, only Rosetta methods have terms describing the membrane protein system. $^a$ High resolution. $^b$ Low resolution. $^c$ The model is tested with 5-fold cross validation over the ``M223" data set. $^d$ The multi-task model is trained with an auxiliary task of globular protein prediction using the ``S2648" data set and is tested on 5-fold cross validation over the ``M223" data set.}
\label{tab:MemMutationPerformance}
\end{table}

\section{Discussion}
 
The adoption of convolutional neural network concept in this work is motivated by the underlying continuity along the distance scale (filtration) dimension. However, unlike images or videos, there is no obvious transferable property of the convolution filters along the convolution dimension in the proposed method, where properties that reside in different distance scales are heterogeneous. To look into this concern, the convolution layers are substituted with ``locally connected layers", where the local connection properties are conserved while the filters applied to different distance scales are allowed to be different. The performance significantly degenerates for the protein-ligand binding affinity prediction example and the task of S350 set prediction in the mutation impact example, which shows that the construction of lower level features in the lower sparse layers benefits from sharing filters along the distance scale and indicates the existence of some common rules for the feature extraction at different distance scales. 

\begin{figure}[ht]
\begin{center}
\includegraphics[keepaspectratio,width=3.3in]{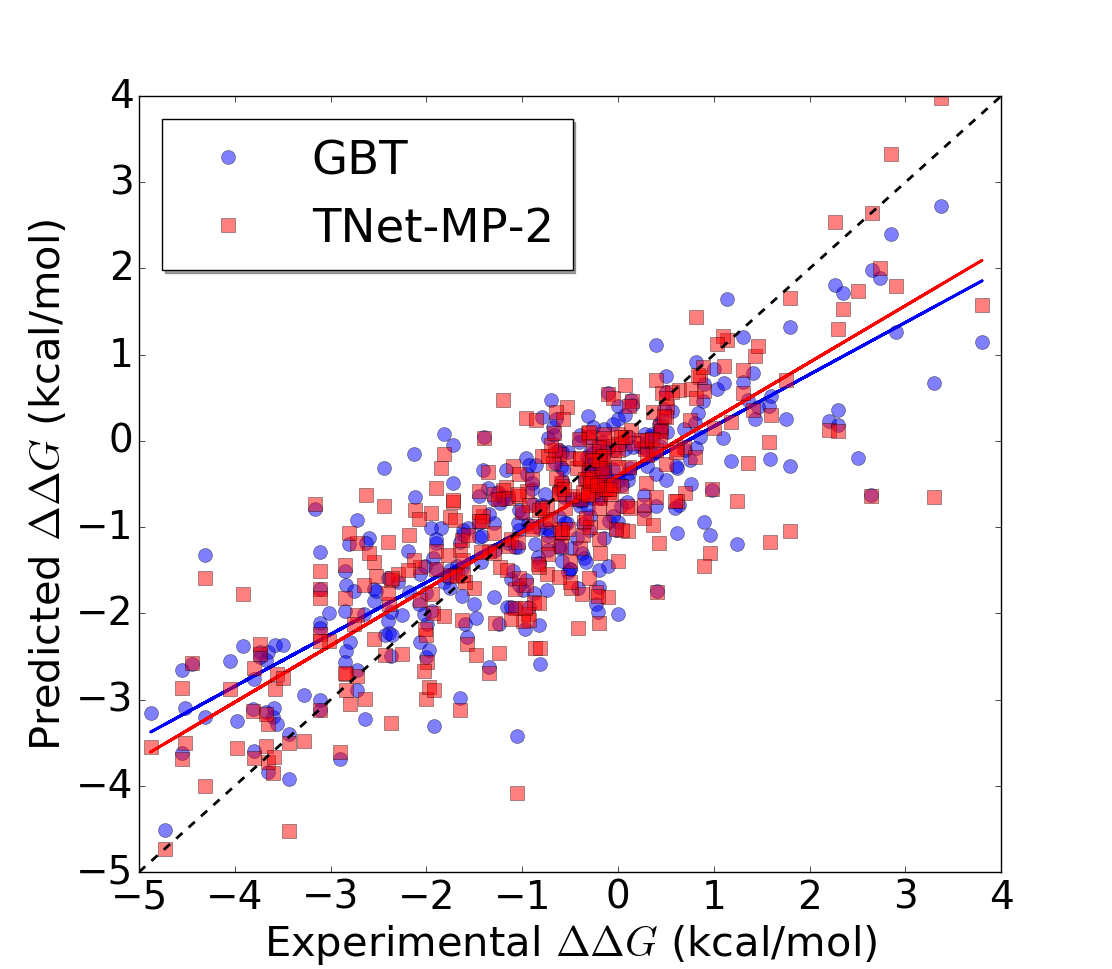}
\caption{A comparison of behaviors of the gradient boosted trees method \cite{ZXCang:2017a} and the neural network based method presented in this work. The plot is for the prediction task of the S350 dataset. The linear fit for GBT prediction is $y = 0.603x - 0.435$ and for TNet-MP-2, $y = 0.657x-0.422$.}
\label{fig:ExtrapolationAbility}
\end{center}
\end{figure}

It is intuitive that the dimension $0$ inputs characterize the pairwise atomic interactions, which clearly contributes to the prediction of energy changes. While it remains unclear whether and to what extent the higher dimensions help with the characterization of the biomolecules, the dimension $0$ inputs and  higher dimensional inputs of Betti numbers  are isolated and tested separately for the task of S350 set in the protein mutation prediction and the protein-ligand binding affinity prediction. To compare the performance of different sets of features, 50 single models are trained for each feature set. Bagging of 20 models from the 50 trained models are randomly generated and repeated 100 times with the median results reported. The individual performances measured by Pearson correlation coefficient (RMSE) for dimension $0$ features are 0.73 (1.09) and 0.82 (1.40) and for dimension $1$ and $2$ features, 0.66 (1.21) and 0.78 (1.54). The combination of all dimensions results in a better performance of 0.74 (1.08) and 0.83 (1.37) showing that the two sets of features both contribute to the prediction and neither is redundant.

Another popular class of machine learning methods is the ensemble of trees methods. Many state-of-the-art methods for biomolecular property prediction are based on random forest (RF) and gradient boosted trees (GBTs). The ensemble of decision trees has the capability of learning complicated functions, but GBTs learn to partition the feature space based on the training data which means that they do not have the ability of appropriate extrapolation of the learned function to broader situations than the training data. Additionally, it is ubiquitous that  data samples are unevenly distributed. It has been observed that in many applications where there are just a handful of samples with large absolute value for the target property, methods of ensemble of trees  tend to overestimate (underestimate) the boarder cases with very negative (positive) target values. In the case of neural network, due to its different way of learning the underlying function, it seems to be able to deliver better results for the boarder cases. Therefore, similar to the idea of bagging, methods of ensemble of trees and neural network based methods may result in different error characteristics for different samples and can potentially improve the prediction power by correcting each others' error when the results from different models are averaged. In the example of prediction of the S350 set, we obtained performance of 0.82 (0.92) for Pearson correlation coefficient (RMSE in kcal/mol) in our other work using handcrafted features with gradient boosted trees \cite{ZXCang:2017a}. When the results are averaged for the two methods, the performance is improved to 0.83 (0.89) which is better than both individual methods. Similar improvement is observed for the protein-ligand binding example. Our method based on handcrafted features and gradient boosted trees with performance 0.82 (1.40) \cite{ZXCang:2017b} and method presented in this work with performance 0.83 (1.37) can achieve improved performance of 0.84 (1.35) when combined by averaging the results. An intuitive illustration is shown in Fig. \ref{fig:ExtrapolationAbility} . It can be visually seen from the plot that the neural network based method presented in this work performs better than the GBT based method for samples with high $\Delta\Delta G$ or with low $\Delta\Delta G$. The slope of linear fitting of the predicted values to the experimental data is 0.66 for the neural network based method and 0.60 for the GBT based method which also illustrates that the neural network based method handles boarder cases better. The observed improvement is marginal since it is mainly on a small portion of the samples.


The approach introduced in this paper utilizes element specific persistent homology to efficiently and sufficiently characterize the 3D biomolecule structures. Convolutional neural network facilitates the automatic feature extraction from raw inputs generated by persistent homology computation. The flexible and hierarchical structure of neural network allows seamless combination of automatically extracted  features and handcrafted features  and also makes it easy to implement multitask learning by combining related tasks to a desired level of model sharing by tuning the layer of model branching. The proposed method can be extended to other applications in the structural prediction of biomolecular properties and has the potential to further benefit from the fast accumulating biomolecular data.

\section*{Acknowledgments}

This work was supported in part by NSF Grants DMS-1160352 and IIS-1302285     and
MSU Center for Mathematical Molecular Biosciences Initiative.

\bibliographystyle{ieeetr}
\bibliography{refs}

\end{document}